\documentclass[twoside,11pt]{article}
\usepackage{jmlr2e}

\usepackage{times}
\usepackage{helvet}
\usepackage{courier}
\usepackage{graphicx}
\usepackage{balance}  
\usepackage{times}
\usepackage{algorithmic}
\usepackage{algorithm}
\usepackage{epsfig}
\usepackage{amssymb}
\usepackage{amsmath}
\usepackage{amsfonts}
\usepackage{subfigure}
\usepackage{nccmath} 
\usepackage{graphicx}
\usepackage{url}
\usepackage{multirow}
\usepackage{wrapfig}
\usepackage{subfigure}
\usepackage{balance}
\usepackage{sidecap}
 \usepackage{sidecap}
 
\usepackage{url}

\begin{document}

\title{Modeling Transitivity in Complex Networks}

\author{\name Morteza Haghir Chehreghani \email morteza.chehreghani@xrce.xerox.com \\
       \addr Xerox Research Centre Europe - XRCE\\
       France
       \AND
       \name Mostafa Haghir Chehreghani \email mostafa.haghirchehreghani@cs.kuleuven.be \\
       \addr Informatics Section\\
       KU Leuven\\
       Belgium}

\editor{--}

\maketitle

\begin{abstract}
An important source of high clustering coefficient in real-world networks is \textit{transitivity}.
However, existing algorithms which model transitivity suffer from at least one of the following problems:
\textit{i}) they produce graphs of a specific class like bipartite graphs,
\textit{ii}) they do not give an analytical argument for the high clustering coefficient of the model, and
\textit{iii}) their clustering coefficient is still significantly lower than real-world networks.
In this paper, we propose a new model for complex networks which
is based on adding transitivity to scale-free models.
We theoretically analyze the model and provide analytical arguments for its different properties.
In particular, we calculate a lower bound on the clustering coefficient of the model which is independent of the network size, as seen in real-world networks.
More than theoretical analysis, the main properties of the model are evaluated empirically and
it is shown that the model can precisely simulate real-world networks from different domains with and different specifications.

\end{abstract}

\section{Introduction}
\label{introduction:sec}
Most of real-world networks such as World Wide Web, social networks, Internet and biological networks
exhibit structural properties which are not in either entirely regular or purely random graphs.
For example, graphs produced by the model of Paul Erd\H{o}s and Alfr\'ed R\'enyi
(the ER model) \cite{jrnl:Erdos}, do not have the two important properties observed in many real-world networks.
The first property is related to the \textit{degree distribution}.
In a network, the \textit{degree distribution} is defined as the probability distribution of the degrees of vertices over the whole network.
In many real-world networks a \textit{power-law distribution} is observed.
More formally, the probability that the degree of a vertex is $k$ is proportional to $k^{- \gamma}$.
Networks with this property are called \textit{scale-free} networks.
However, the degree distribution of graphs produced by the ER model converges to a \textit{Poisson distribution}.

The second property is related to the \textit{clustering coefficient}.
Clustering coefficient is used to measure how well vertices in a network tend to be clustered together.
In most of real-world networks, vertices tend to create tight groups characterized by dense ties \cite{jrnl:Watts}.
However, in the ER model, every two vertices are connected with a constant and independent probability and therefore, the model generates graphs with a low clustering coefficient.

The $\beta$ model (the Watts-Strogatz model), proposed by Watts and Strogatz \cite{jrnl:Watts},
produces graphs with the \textit{small-world} property and high \textit{clustering coefficient}.
In small-world networks, the distance between each pair of vertices is proportional to the logarithm of the number of vertices in the network.
However, the $\beta$ model produces an unrealistic degree distribution.
The Barab\'{a}si-Albert (BA) model, proposed by Albert-L\'{a}szl\'{o} Barab\'{a}si and R\'{e}ka Albert
produces \textit{scale-free} graphs \cite{jrnl:Barabasi}.
The model is based on two important concepts: \textit{growth} and \textit{preferential attachment}.
\textit{Growth} means that the number of vertices in the network increases over time.
\textit{Preferential attachment} means that vertices with higher degree are more likely to receive new edges.
The degree distribution of a graph resulting from the BA model is a power-law in the form of
$\Pr[k] \sim k^{-3}$.
However, the clustering coefficient of graphs produced by the BA model is
significantly lower than the clustering coefficient of real-world networks. 
%
%
Takemoto and Oosawa \cite{jrnl:Takemoto} propose a model for evolving networks by merging complete graphs (cliques) as building blocks.
The model shows power-law degree distribution, power-law clustering spectra and high average clustering coefficients independent of the size of network.
However, in most cases, real-world networks are evolved in a different way:
they usually \textit{grow} during the time by obtaining new vertices,
rather than by merging complete graphs.

An important source of high clustering coefficient in networks is \textit{transitivity}.
Transitivity means if $u$ is connected to $v$ and $v$ is connected to $w$, 
the probability of having a connection between $u$ and $w$ is higher than any other pair of vertices in the network.
Most of edges in real-world networks are \textit{local} and they are drawn between vertices which have a common neighbor \cite{conf:Leskovec}.
The model of \cite{jrnl:Newman2} incorporates transitivity and generates graphs with high clustering coefficient.
However, it produces bipartite networks which are limited to situations like company directors and movie actors.
Clustering coefficient in the graphs produced by the model of \cite{jrnl:Li} is still significantly  
lower than clustering coefficient of real-world networks. 
Leskovec et.al. \cite{conf:Leskovec} propose several mechanisms for modeling transitivity in complex networks.
However, they do not provide any theoretical argument for the clustering coefficient of the mechanisms. 
The importance of such a theoretical analysis is that
it guarantees that the model will reflect important properties of real-world networks,
since a high clustering coefficent, independent of the network size, is seen in many real-world networks.  
On the other hand, for most of network models, it is not easy to theoretically analyze the clustering coefficent.
For example, up to now, clustering coefficient of BA networks has only been determined by
numerical simulations\footnote{Numerical simulations show that clustering coefficent of a BA network with $n$ vertices is $n^{-0.75}$.},
and it is known to be very difficult to theoretically analyze it.
Therefore, it is interesting to develop a model for transitivity in complex networks
such that its clustering coefficent can be verified by theoretical arguments.  


In this paper, we present the $\eta$ model for modeling transitivity in complex networks.
At every time interval $t$, the network obtains a new vertex and
the new vertex is connected to some existing vertices. This step is similar to the BA model.
Then, each vertex is selected with a probability proportional to its \textit{degree} .
If it is selected, then a pair of its neighbors are chosen randomly and an edge is drawn between them.
The model has two adjustment parameter $\eta$ and $m$. 
We theoretically analyze the model and prove that it produces networks with power-law degree distribution,
high clustering coefficient and the small-world property.
Compared to the clustering coefficient of random graphs or graphs produced by existing scale-free models,
the clustering coefficient of the $\eta$ model is significantly higher.
In particular, by theoretical arguments, we prove that it is independent of the network size
and depends solely on parameters like $\eta$ and $m$.
We also empirically evaluate the model and
show that it can precisely simulate networks from different domains (biology, technology, social and information networks) with different characteristics. 
The rest of this paper is organized as follows.
In the second section, we present the model and theoretically analyze its important properties.
In the third section, we empirically evaluate the model and
show that it can produce graphs very close to real-world networks.
An overview of related work is given in Section \ref{sec:relatedwork}.
%
\section{The $\eta$ model}
\label{model:sec}


In this section, we first present the $\eta$ model and then
we theoretically analyze its important properties like power-law degree distribution, high clustering coefficient and the small-world property.


Algorithm \ref{alg:model} describes the high level pseudo code of the $\eta$ model proposed for modeling transitivity in complex networks.
First a small graph $\mathbf G_0$ is produced.
We refer to it as the \textit{initial graph}.
Then, at every time interval $t \in \{1,\hdots,\mathbf T \}$,
the following steps are performed:\\
I. \textit{growth.}
A new vertex $v$ is added to the network $\mathbf G$.
We denote by $t_v$ the time of adding $v$ to $\mathbf G$.\\
II. \textit{preferential attachment.}
The vertex $v$ is connected to $m$ existing vertices.
Existing vertices are chosen based on their degree.
While every model which produces scale-free networks can be used, for the sake of simplicity,
we here use the basic BA model.
Therefore, for $m$ times, a vertex $w$ with probability
\begin{equation}
\frac{d_w(t)}{2e(t)}
\end{equation}
is chosen and connected to $v$.
We denote by $d_w(t)$ the degree of $w$ at time interval $t$ and by $e(t)$ the number of edges of the graph at time interval $t$.\\
III. \textit{transitivity.}
At this step each vertex $w$ of the graph is selected with probability
\begin{equation}
\frac{\eta d_w(t)}{2e(t)}
\end{equation}
where $\eta$ is a non-negative real number.
Then, if $w$ is selected, among the neighbors of $w$, two vertices are chosen \textit{uniformly at random} and are connected to each other.
%
\begin{algorithm}
\caption{High level pseudo code of the $\eta$ model.}
\label{alg:model}
\mbox{\textsc{GraphGenerator}}
\begin{algorithmic} [1]
\REQUIRE {A non-negative real number $\eta$, a non-negative integer $\mathbf T$, a non-negative integer $m$}.
\ENSURE A graph $\mathbf G$ generated by the $\eta$ model.

\STATE initialize $\mathbf G$ by a small graph 

\FOR {$t=1$ \textbf{to} $\mathbf T$}

\STATE\COMMENT{\textit{growth}:}
\STATE {add a new vertex $v$ to $\mathbf G$}

\STATE\COMMENT{\textit{preferential attachment}:}
\STATE {connect $v$ to $m$ existing vertices}
\COMMENT{every existing vertex is selected proportional to its degree}

\STATE\COMMENT{\textit{transitivity}:}
\FOR {every vertex $w \in V_{\mathbf G}$}
\STATE select $w$ with probability $\frac{\eta d_w(t)}{2e(t)}$
\IF {$w$ is selected}
\STATE select two neighbors $x$ and $y$ of $w$ uniformly at random
\STATE draw an edge between $x$ and $y$
\ENDIF
\ENDFOR
\ENDFOR
\RETURN $\mathbf G$
\end{algorithmic}
\end{algorithm}



The authors of \cite{conf:Leskovec} investigated different cases of producing triangles in complex networks.
In their scenario, a source vertex $u$ decides to connect to some vertex $w$ whose distance with $u$ is two.
$u$ first selects a neighbor $v$ and then $v$ selects a neighbor $w \neq u$.
$u$ and $v$ might use different policies to select $v$ and $w$, e.g. uniform selection or selecting based on degree.
Here we first select $v$ proportional to its degree and then, $u$ and $w$ are selected uniformly at random. 
The main contribution of this work compared to \cite{conf:Leskovec} is that we precisely formulate the procedure,
which gives us a possibility to analytically study the model.
Particularly, we provide a lower bound on the clustering coefficient independent of the network size.

\subsection{Expected number of edges}
\label{sec:edges}

In this section, we calculate the expected number of edges of the network at every time interval $t$.

The number of edges at time interval $t$, i.e. $e(t)$, satisfies the following dynamical equation:
\begin{align}
\frac{\partial e(t)}{\partial t} &= \underbrace{ m }_\text{preferential attachment}  + \underbrace{ \sum_{w\in V_{\mathbf G}(t)}\frac{\eta d_{w}(t)}{2e(t)} }_\text{transitivity} &= m+\eta \nonumber
\end{align}
where $V_{\mathbf G}(t)$ denotes vertices of $\mathbf G$ at time interval $t$.
After solving this equation, we obtain
\begin{align}
\label{eq:edge}
e(t) = (m+\eta)t + e(\mathbf G_0) 
\end{align}
where $e(\mathbf G_0) $ denotes the number of edges in the initial graph.
For large enough $t$, we sometimes ignore $e(\mathbf G_0) $ and consider $e(t)$ as $(m+\eta)t $.

\subsection{Power-law degree distribution}
\label{sec:powerlaw}

In this section, we show that
in a graph produced by the $\eta$ model,
vertices (except those added at the very early time intervals) have a power-law degree distribution.



At every time interval $t \in \{1,\hdots, \mathbf T \}$,
every vertex $v$ in the network satisfies the following dynamical equation:
\begin{align}
\frac{ \partial d_v(t)}{ \partial t} &\overset{(a)}{\approx}\underbrace{ \sum_{u \in N_v(t)} \left(\frac{\eta d_u(t)}{2 e(t)} \times \frac{2}{d_u(t)} \right) }_\text{transitivity} + \underbrace{ \frac{md_v(t)}{2e(t)} }_\text{preferential attachment} \nonumber \\ 
                                                       &= \sum_{u \in N_v(t)} \left(\frac{\eta}{ e(t)}\right) + \frac{md_v(t)}{2e(t)} = \frac{\eta d_v(t)}{ e(t)} + \frac{m d_v(t)}{2e(t)} \label{eq:new12}
\end{align}
where $N_v(t)$ refers to neighbors of vertex $v$ at time interval $t$.

The approximation ${(a)}$ is employed to make the computation of the  dynamical equation $\frac{\partial d_v(t)}{\partial t}$ feasible, since, otherwise it would require taking the expectation of a function with the random variable at the denominator (i.e. the number of edges), which is computationally intractable. In principle, one could use the polynomial normal forms of such functions to eliminate the denominator. However, this transformation yields an exponential order in the number of conjunctions. Therefore, in mean-field theory, it is proposed to approximate the expectation via replacing the random denominator by its expectation, i.e. by $\mathbb E[f/g]\approx f/\mathbb E[g]$, where $f$ is nonrandom~\cite{jrnl:GKR01,jrnl:HB97}.
This approximation is exact in the thermodynamic limit, i.e. for large enough $t$, for example when $t>20$.
One can obtain higher order improvements of the approximation e.g. by a Taylor expansion around the expectation.  The quality of such an approximation has been investigated in the context of mean-field theory by Markov Chain Monte Carlo (MCMC) simulations. Based on extensive experimental
evidences for example in~\cite{jrnl:HB97,jrnl:PHB00}, the first-order approximation competes with more refined techniques such as the TAP method~\cite{jrnl:TAP77}. Moreover,  for large enough $t$, as mentioned earlier, the approximation becomes almost exact and the higher order approximation terms diminish.~\footnote{In our MCMC simulations with $1,000$ runs, the approximation is unbiased, i.e. the difference between the mean of the empirical distribution and the approximated quantity is only $0.061$ times the standard deviation.}

By replacing $e(t)$ with the value obtained in Equation \ref{eq:edge}, 
for large enough $t$, Equation \ref{eq:new12} amounts to
\begin{align}
\frac{ \partial d_v(t)}{ \partial t} &= \frac{2 \eta + m}{2(\eta + m)} \times \frac{d_v(t)}{t} =  \frac{ \alpha d_v(t)}{t} \label{eq:new2}
\end{align}
where $\alpha=\frac{2 \eta + m}{2(\eta + m)}\,$.
%

To solve Equation \ref{eq:new2}, we need to find the initial degree of vertex $v$, i.e. the number of edges $v$ finds when it is added to the network at $t_v$.
At time interval $t_v$, $v$ finds $m$ edges due to preferential attachment, and 
it expects to find $\frac{\eta m}{e(t_v)}$ edges due to transitivity.
Therefore, its initial degree will be $m+\frac{\eta m}{(m+\eta)t_v}$.

Then, using the \textit{continuum theory} \cite{jrnl:Albert}, we obtain
\begin{equation}
d_v(t) = \left( m+\frac{\eta m}{(m+\eta)t_v} \right)  \left(\frac{t}{t_v}\right)^{\alpha} \, ,
\label{eq:powerlaw1}
\end{equation}
particularly
\begin{equation}
d_v(\mathbf T) = \left( m+\frac{\eta m}{(m+\eta)t_v} \right) \left(\frac{\mathbf T}{t_v}\right)^{\alpha} .
\label{eq:powerlaw2}
\end{equation}

%

If $v$ is added to the network at a large enough time interval (i.e. $t_v$ is larger than a lower bound $L$),
Equations \ref{eq:powerlaw1} and \ref{eq:powerlaw2} can be written as
\begin{equation}
d_v(t) = m \left(\frac{t}{t_v}\right)^{\alpha} 
\label{eq:powerlaw3}
\end{equation}
and
\begin{equation}
d_v(\mathbf T) = m  \left(\frac{\mathbf T}{t_v}\right)^{\alpha} 
\label{eq:powerlaw4}
\end{equation}

The probability that at time interval $\mathbf T$ a vertex $v$ has a degree $d_v(\mathbf T)$ smaller than $k$ is
\begin{align}
\Pr[d_v(\mathbf T) <  k] &= \Pr[ m \left( \frac{\mathbf T}{t_v} \right)^{\alpha} < k ] = \Pr[ t_v > \frac{\mathbf T  \times m^{\frac{1}{\alpha}} } {k ^{\frac{1}{\alpha}}} ]
\end{align}

and
\begin{equation}
\Pr[d_v(\mathbf T) <  k] = 1- \Pr[ t_v \leq \frac{\mathbf T \times m^{\frac{1}{\alpha}} } {k ^{\frac{1}{\alpha}}} ]
\label{eq:prpbability}
\end{equation}

We suppose that the vertices are added to the network at equal time intervals
$
\Pr[t_v] = \frac{1}{\mathbf T} 
$. Putting it into Equation \ref{eq:prpbability}, we get
\begin{align}
\Pr[d_v(\mathbf T) < k] &= 1 - \frac{\mathbf T \times m^{\frac{1}{\alpha}} } {\mathbf T \times k ^{\frac{1}{\alpha}}} = 1 - \left( \frac{m}{k} \right)^{\frac{1}{\alpha}}
\end{align}

Then, the degree distribution $\Pr [k]$ can be computed as
\begin{align}
\Pr[k] &= \frac{\partial \Pr[d_v(\mathbf T)<k]}{\partial k} = \frac{m^{\frac{1}{\alpha}} }{\alpha} \times k^{-(1+\frac{1}{\alpha})}
\end{align}

which means $\Pr[k] \sim  k^{-(1+\frac{1}{\alpha})}$.
Therefore, we have a power law degree distribution $\Pr[k] \sim k^{-\gamma}$,
where
\begin{align}
\gamma &= 1+\frac{1}{\alpha}      = \frac{4\eta+3m}{2\eta+m}  =  2 + \frac{m}{2\eta+m}    \label{eq:gamma}
\end{align}

\subsection{The small world property}
\label{sec:smallworld}

Reuven Cohen and Shlomo Havlin \cite{jrnl:Cohen} showed that scale-free networks have a small diameter.
Particularly, using analytical arguments, they showed that scale-free networks with $2<\gamma<3$
have a very small diameter proportional to $\mathbf{ln} \mathbf{ln} n$.
They also showed that for $\gamma=3$ the diameter is proportional to $\frac{\mathbf{ln} n}{\mathbf{ln} \mathbf{ln} n}$,
and for $\gamma>3$ it is proportional to $\mathbf{ln} n$.
In all cases the scale-free network satisfies the small-world property.
We note that here the diameter is the mean distance between vertices.
As Equation \ref{eq:gamma} indicates, for the $\eta$ model we have: $2 \leq \gamma  \leq 3$.
Particularly, for non-zero values of $\eta$ and $m$, we have $2 < \gamma < 3$.
This means that the $\eta$ model satisfies the required conditions, i.e. it produces graphs with the small-world property where the diameter is proportional to $\mathbf{ln} n$.

\subsection{Clustering coefficient}
\label{sec:clusteringcoefficient}

In this section, 
we provide a lower bound on the clustering coefficient of the networks produced by the $\eta$ model,
which is independent of the network size and depends only on the $\eta$ and $m$ parameters. 

Watts and Strogatz \cite{jrnl:Watts} defined the clustering coefficient
of a network as\footnote{An alternative definition of the clustering coefficient,
also widely used, was introduced by Barrat and Weigt \cite{jrnl:Barrat}:
$\frac{3 \times \text{number of triangles in the network}}{\text{number of connected triples of vertices}}$.}
\begin{equation}
\label{eq:clusteringcoef}
\left<CC\right>=\frac{1}{n}\sum_{v \in V_{\mathbf G}} \left<CC_v\right>
\end{equation}
where $n$ is the number of vertices of the network and

\begin{equation}
\left<CC_v\right>=\frac{C_v}{ \binom{d_v}{2} }
\end{equation}
where $C_v$ is the number of edges among the neighbors of $v$.
$\left<CC_v\right>$ is called the local clustering coefficient of $v$.

For a network produced by the $\eta$ model, $C_v$ can be written as
\begin{equation}
C_v=\sum_{t=t_v}^{\mathbf T} \left( \left<C_v\right>_T(t) + \left<C_v\right>_P(t) \right)
\end{equation}
where
\begin{itemize}
\item $\left<C_v\right>_P(t)$ is the number of edges between neighbors of $v$
which are added to $\mathbf G$ during the \textit{preferential attachment} step at time interval $t$, and
\item $\left<C_v\right>_T$ is the number of edges between neighbors of $v$
which are added to $\mathbf G$ during the \textit{transitivity} step at time interval $t$.
\end{itemize}

Then, for a vertex $v$, at every time interval $t\geq t_v$, we define $\tau_v(t)$ as
\begin{equation}
\tau_v(t) = \sum_{t'=t_v}^{t}  \left<C_v\right>_T(t')
\end{equation}

We have

\begin{equation}
C_v \geq \tau_v(\mathbf T)
\end{equation}

Therefore 
\begin{equation}
\label{eq:CC_v1}
\left<CC_v\right> \geq \frac{\tau_v(\mathbf T)}{\binom{d_v(\mathbf T)}{2}} 
\end{equation}

%
%
%
%

Suppose that $v$ is added to the network at a time interval greater than a lower bound $L$ (i.e. $t_v\geq L$)
such that we can use Equation \ref{eq:powerlaw3} to describe its degree.
In the following, we compute $\tau_v(\mathbf T)$.

For $t\geq t_v$, $\tau_v$ satisfies the dynamical equation
\begin{align}
\frac{\partial \tau_v(t)}{\partial t}  &= \frac{\eta  d_v(t)}{2e(t)}  = \frac{\eta m t^{\alpha-1}}{2(\eta+m){t_v}^{\alpha}} \label{eq:cc1}
\end{align}


Then, at time interval $\mathbf{T}$, we will have:
\begin{align}
\label{eq:int1}
\tau_v(\mathbf T) - \tau_v(t_v) &=\int_{t_v}^\mathbf{T} \frac{\eta m }{2(m+\eta) {t_v}^{\alpha}} \times t^{\alpha-1} \partial t  
\end{align}

To solve this dynamical equation, we need to find $\tau_v(t_v)$.
Since at time interval $t_v$ vertex $v$ finds $m+\frac{\eta m}{(m+\eta)t_v}$ edges,
$\tau_v(t_v)$ will be:
\begin{align}
\tau_v(t_v)  =     \frac{\eta \times \left( m+\frac{\eta+m}{(m+\eta)t_v} \right)}{2(m+\eta)t_v}  \geq     \frac{\eta m }{2(m+\eta)t_v} 
\end{align}

Therefore after solving the integral of Equation \ref{eq:int1}, we will have
\begin{align}
\tau_v(\mathbf T)  &\geq    m K \times \left(  \frac{ \mathbf{T}^{\alpha} } {2{t_v}^{\alpha}} - \frac{1}{2} \right) + \frac{\eta m}{2(m+\eta)t_v} \geq  \frac{ mK \mathbf{T}^{\alpha} } {2{t_v}^{\alpha}} - \frac{m K}{2} 
\end{align}
where $K=\frac{\eta }{\alpha \times(m+\eta)} = \frac{2\eta}{2\eta+m}$ .

%

Now, we use Equation \ref{eq:CC_v1} to find a lower bound for $\left<CC_v\right> $:

\begin{align}
\left<CC_v\right>  &\geq  \frac{\tau_v(\mathbf T)}{\binom{d_v(\mathbf T)}{2}} \geq  \frac{ 2\tau_v(\mathbf T)}{d_v(\mathbf T)^2}  \geq \frac{ K {t_v}^{\alpha}}{m{\mathbf{T}^{\alpha}}} - \frac{ K {t_v}^{2\alpha}}{m {\mathbf{T}^{2\alpha}}} \label{eq:CCbound}
\end{align}

Let $v$ be a vertex such that $ L \leq t_v \leq \mathbf T $.
Up to now, we have computed a lower bound for $\left<CC_v\right> $.
Now, we want to compute a lower bound for the clustering coefficient of the network induced by the vertices added to the network at time intervals $t_L,t_{L+1},\hdots,t_{\mathbf T}$.
We refer to this quantity as $\left<CC\right> $ since it is almost the clustering coefficient of the whole network
(compared to $\mathbf T$, $L$ is very small and for only a few vertices we can not use Equation \ref{eq:powerlaw3} to describe the degree).

Using Equations \ref{eq:clusteringcoef} and \ref{eq:CCbound}, we obtain
\begin{align}
\left<CC\right> &\geq \frac{1}{\mathbf{T}-L+1} \sum_{t_v=L}^\mathbf{T} \left(  \frac{ K {t_v}^{\alpha}}{m{\mathbf{T}^{\alpha}}} - \frac{ K {t_v}^{2\alpha}}{m{\mathbf{T}^{2\alpha}}} \right) \label{eq:lowerbound2}
\end{align}

A simple form of the Riemann sum \cite{book:thomas} says: ($r,a,b>0$)
\begin{equation*}
\sum_{x=a}^b {x}^{r} \geq \int_{a-1}^ b {x}^{r} \partial x
\end{equation*}

This inequality and Equation \ref{eq:lowerbound2} yield
\begin{align}
\left<CC\right>  &\geq \frac{1}{\mathbf{T}-L+1} \int_{L-1}^\mathbf{T}  \left(  \frac{ K {t_v}^{\alpha}}{m{\mathbf{T}^{\alpha}}} - \frac{ K {t_v}^{2\alpha}}{m {\mathbf{T}^{2\alpha}}} \right) \partial t_v \label{eq:lowerbound3}
\end{align}

After solving the integral, we obtain

\begin{align}
\left<CC\right>  \geq \frac{K}{m(\alpha+1)} - \frac{K}{m(2\alpha+1)} = \frac{2\eta (\eta+m)}{m(4\eta+3m)(3\eta+2m)} \label{eq:finallowerbound}
\end{align}
%
%
%
%

Therefore, a lower bound is provided for the clustering coefficient of a $\eta$ network,
which is independent of the network size and depends on the $\eta$ and $m$ parameters.
We refer to Equation \ref{eq:finallowerbound} as $B$.

\subsection{Simulating real-world networks}
\label{sec:realworldsimulation}

\begin{table*}
\caption{\label{table:networkproperties}Real-world networks and the equivalent networks produced by the $\eta$ model.
$\mathcal C$ and $\mathcal C_{\eta}$ are
clustering coefficient of the real-world networks and
clustering coefficient of the networks produced by the $\eta$ model, respectively.
$\mathcal C_{BA}$ is the clustering coefficient of the simulated network if transitivity is not used.
}
\centering
\begin{tabular}{|l l l l |l l l l l |}
\hline
\multicolumn{4}{|c}{\multirow{1}{*}{\textbf{Real-world networks}}} & \multicolumn{5}{|c|}{\textbf{Simulated networks}}             \\
Network & \# vertices & \# edges & $\mathcal C$  & $m$ & $\eta$ & \# edges & $\mathcal C_{\eta}$ & $\mathcal C_{BA}$ \\
\hline
electronic circuits   & $24,097$  & $53,248$ & $0.03$ & $2$ & $0.23$  & $53,121 $ & $0.034$ & $0.009$ \\
email address books   & $16,881$  & $57,029$ & $0.13$ & $3$ & $0.5$   & $58,041$ & $0.11 $  & $ 0.0047$ \\
marine food web       & $135$     & $598$    & $0.23$ & $4$ & $0.54$ & $599$   & $0.24$ & $0.148$ \\
neural network        & $307$     & $2,359$  & $0.28$ & $5$ & $2.8$   & $2,341$  & $0.29$ & $0.098$ \\
Roget's thesaurus     & $1,022$   & $5,103$  & $0.15$ & $4$ & $1.4$  & $5,389$  & $0.14$ & $0.038$  \\
\hline
\end{tabular}
\end{table*}

In this section, we consider several real-world networks, with different specifications and from different domains including 
biology, technology, social and information networks,
and try to simulate them using the $\eta$ model.
Table \ref{table:networkproperties} summarizes the characteristics of different real-world networks and the networks simulating them.
Note that we only describe one way of simulating the real-world networks by the $\eta$ model
which is not unique and the only existing way.
In all simulated networks, the initial graph simply consists of two vertices connected by an edge.

\begin{figure*}[ht]
\centering
\subfigure[Electronic circuits]
{
\includegraphics[width=0.315\textwidth]{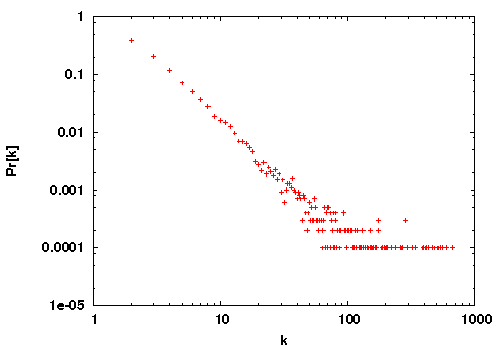}
\label{figure:electronicCircuitsdegree}
}
\subfigure[Email address books]
{
\includegraphics[width=0.315\textwidth]{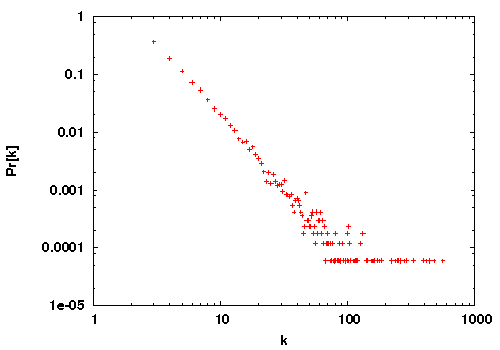}
\label{figure:emailAddressBooksdegree}
}
\subfigure[marine food web]
{
\includegraphics[width=0.305\textwidth]{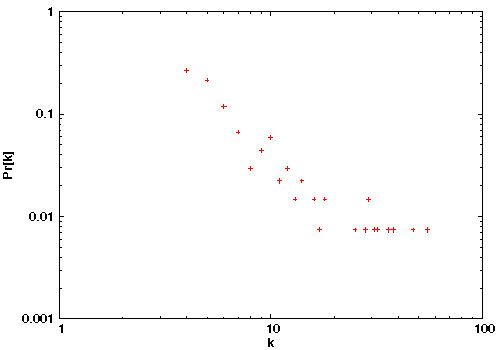}
\label{figure:marineFoodWebdegree}
}
\subfigure[nematode C. Elegans.]
{
\includegraphics[width=0.315\textwidth]{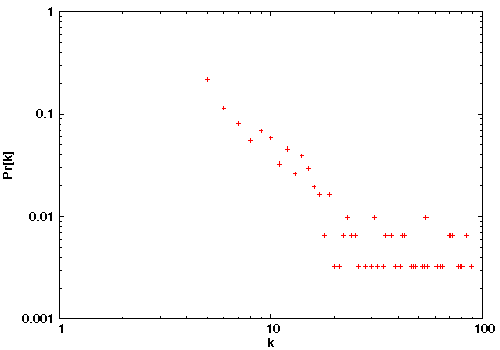}
\label{figure:neuralNetworkdegree}
}
\subfigure[Roget's thesaurus]
{
\includegraphics[width=0.315\textwidth]{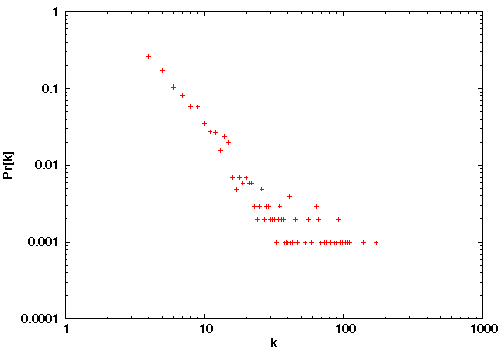}
\label{figure:Rogetdegree}
}
\caption
{
\label{figure:degreedistribution}
Degree distribution of the $\eta$ networks produced for different real-world networks.}
\end{figure*}

The first real-world network studied here is the \textit{electronic circuits} network.
In this network vertices are electronic components e.g., logic gates in digital circuits and resistors,
capacitors and diodes in analogic circuits and edges are the wires \cite{jrnl:Cancho}.
It has $24,097$ vertices and $53,248$ edges and its clustering coefficient is $0.030$.
In order to simulate this network, we produce an $\eta$ graph with these parameters:
$m=2 $ and $\eta=0.23 $ and it has the same number of vertices as the electronic circuits network.
The graph produced by the $\eta$ model has $53,121$ edges, its clustering coefficient is $0.034$
and its degree distribution is depicted in Figure \ref{figure:electronicCircuitsdegree}.

The second real-world network is the network of \textit{email address books} \cite{jrnl:NewmanEmail}.
In this network, vertices represent computer users and 
an edge is drawn from user A to user B if B's email address appears in A's address book.
This network has $16,881$ vertices and $57,029$ edges and its clustering coefficient is $0.13$.
We simulate this network by the $\eta$ model using the following parameters: $m=3$ and $\eta=0.5$
and the number of vertices in the produced graph is $16,881$.
The clustering coefficient of the simulated network is $0.11$.
However, if we remove transitivity from the network (and produce a BA network), its clustering coefficient will be only $0.0047$.
Figure \ref{figure:emailAddressBooksdegree} presents degree distribution of the simulated network.

The next two real-world networks are biological networks.
In the \textit{marine food web} network, vertices represent species
in an ecosystem and an edge from vertex A to vertex B indicates that A preys on B \cite{jrnl:Huxham} and \cite{jrnl:CohenFood}.
This network has $135$ vertices and $598$ edges and its clustering coefficient is $0.23$.
The following parameters are used by the $\eta$ model to simulate this network: $m=4$, $\eta=0.54$, and number of vertices is $135$.
The produced graph has $599$ edges and its clustering coefficient is $0.24$.
Figure \ref{figure:marineFoodWebdegree} presents degree distribution of the networks
simulated by the $\eta$ model.

The other important class of biological networks are \textit{neural networks}.
The neural network of the nematode C. Elegans reconstructed by White et al. \cite{jrnl:White}
has $307$ vertices and  $2,359$ edges and its clustering coefficient is $0.28$.
We simulate it by a $\eta$ network with $m=5$ and $2.8$.
The clustering coefficient of the produced graph is $0.29$.
Degree distribution of the simulated network is shown in Figure \ref{figure:neuralNetworkdegree}.

The last real-world network investigated in this paper is the \textit{Roget's thesaurus} network \cite{book:Knuth}.
Each vertex of the graph corresponds to one of the $1,022$ categories in the $1,879$ edition of Peter Mark Roget's Thesaurus of English Words and Phrases.
An edge is drawn between two categories if Roget gave a reference to the latter among the words and phrases of the former,
or if the two categories were related to each other by their positions in Roget's book.
This network has $5,103$ edges and its clustering coefficient is $0.15$.
We simulate it by a $\eta$ network with $m=4$ and $\eta=1.4$.
The produced graph has $5,389$ edges and its clustering coefficient is $0.14$.
Figure \ref{figure:Rogetdegree} presents degree distribution of the simulated network.

\subsection{Empirical evaluation of the $\eta$ model}
\label{sec:empiricalpropertiesevaluation}

\begin{figure*}[ht]
\centering
\subfigure[The $\eta$ parameter.]
{
\includegraphics[width=0.4\textwidth]{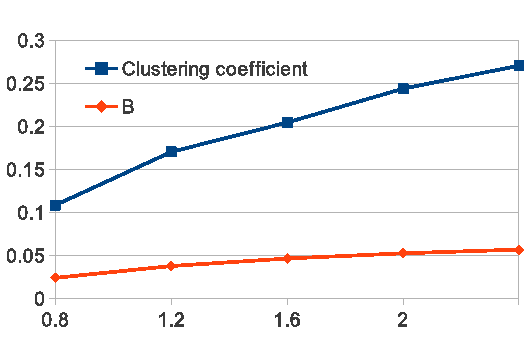}
\label{figure:bound}
}
\subfigure[The $m$ parameter.]
{
\includegraphics[width=0.4\textwidth]{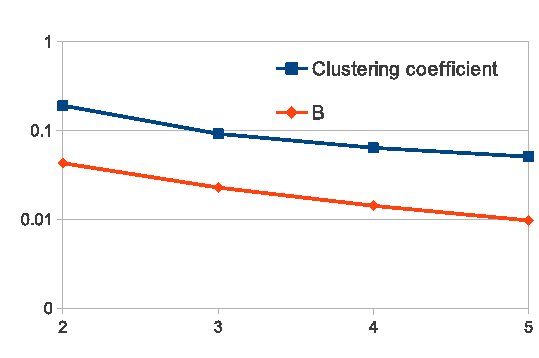}
\label{figure:bound2}
}
\caption
{
\label{figure:degreedistribution2}
Effect of $\eta$ and $m$ on the clustering coefficient and the bound $B$}
\end{figure*}

In this section, we empirically evaluate the different properties of the $\eta$ model.
In order to investigate the impact of $\eta$, we fix $m$ to $2$ and $n$ to $10,000$,
and produce networks with different $\eta $: $0.4$, $0.8$, $1.2$, $1.6$, $2$.
Figure \ref{figure:degreedistribution} illustrates the degree distributions of the produced networks.
If $\eta$ is set to $0$, a BA network is obtained.
As we see in the figure, the degree distributions follow a power-law.
Furthermore, by increasing $\eta$, the exponent $\gamma$ slowly increases
which is consistent with Equation \ref{eq:gamma}.
Figure \ref{figure:bound} compares the clustering coefficient of the networks and the bound $B$ obtained in Equation \ref{eq:finallowerbound}.
In the produced networks, $m$ is $2$, $n$ is $10,000$ and $\eta$ varies between $0.4$ and $2$.
It shows that by increasing the clustering coefficient, $B$ increases as well.
Table \ref{table:dimacc} summarizes the characteristics of the produced networks.
In the produced networks, by increasing $\eta$, the clustering coefficient and average degree increase and the diameter decreases.

\begin{figure*}[htb]
\centering
\subfigure[BA network with $m=2$]
{
\includegraphics[scale=0.33]{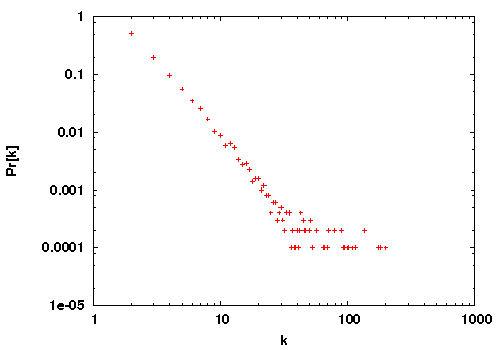}
\label{fig:subfig1}
}
\hspace{-5mm}
\subfigure[$\eta$ model, $m=2$, $\eta=0.8$]
{
\includegraphics[scale=0.33]{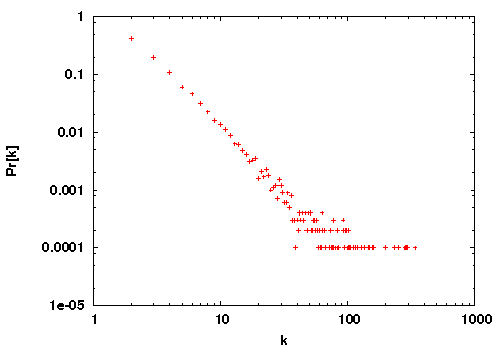}
\label{fig:subfig3}
}
\hspace{-5mm}
\subfigure[$\eta$ model, $m=2$, $\eta=1.6$]
{
\includegraphics[scale=0.33]{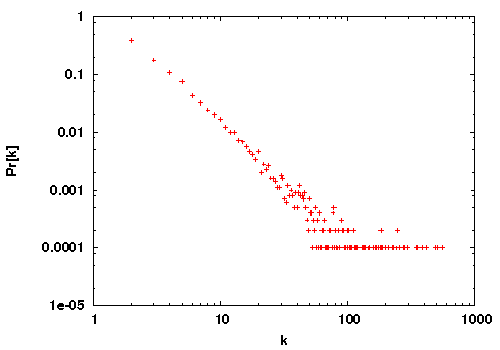}
\label{fig:subfig3}
}
\hspace{-5mm}
\subfigure[$\eta$ model, $m=2$, $\eta=2$]
{
\includegraphics[scale=0.33]{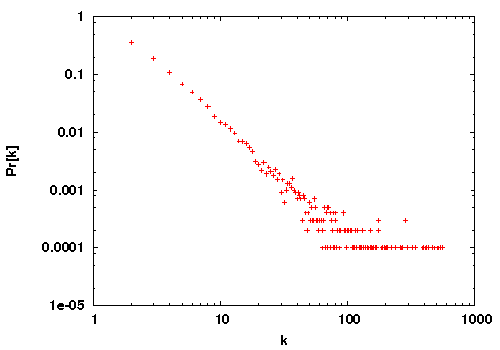}
\label{fig:subfig3}
}
\caption
{
\label{figure:degreedistribution}
Comparison of degree distributions of a BA network and five $\eta$ networks having different values of $\eta$.
}
\end{figure*}


\begin{table}
\caption{\label{table:dimacc}
Diameter, clustering coefficient, and average degree of
networks produced by the $\eta$ model for different values of $\eta$.
$n$ is set to $10,000$ and $m$ is set to $2$.
}
\centering
\begin{tabular}{l l l l}
\hline
$\eta$ & diameter & clustering coefficient & Avg. degree\\
\hline
$0$   & $5.28$ & $0.0045$  & $4 $ \\
$0.4$ & $4.91$ & $0.108$   & $ 4.701$ \\
$0.8$ & $4.72$ & $0.171$   & $ 5.432$ \\
$1.2$ & $4.21$ & $0.204$   & $ 6.149$ \\
$1.6$ & $3.67$ & $0.244$   & $ 6.900$ \\
$2$   & $3.25$ & $0.27$    & $ 7.679$ \\
\hline
\end{tabular}
\end{table}

%
%

\begin{figure*}[htb]
\centering
\subfigure[$m=2, \eta=1$]
{
\includegraphics[scale=0.33]{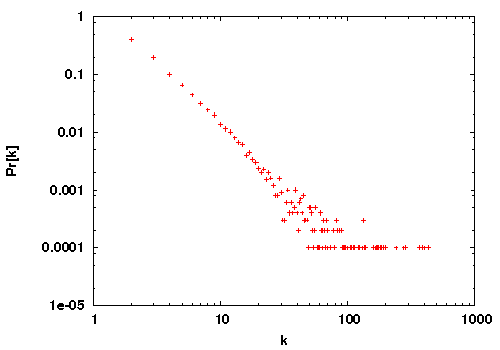}
\label{fig:subfig1}
}
\hspace{-5mm}
\subfigure[$m=3, \eta=1$]
{
\includegraphics[scale=0.33]{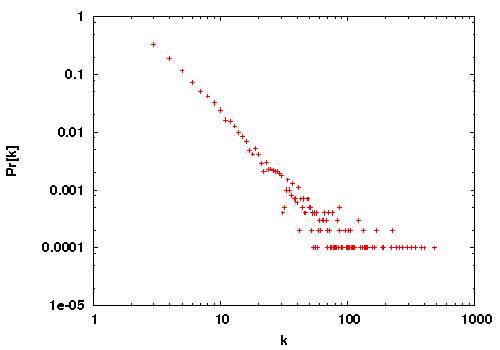}
\label{fig:subfig2}
}
\hspace{-5mm}
\subfigure[$m=4, \eta=1$]
{
\includegraphics[scale=0.33]{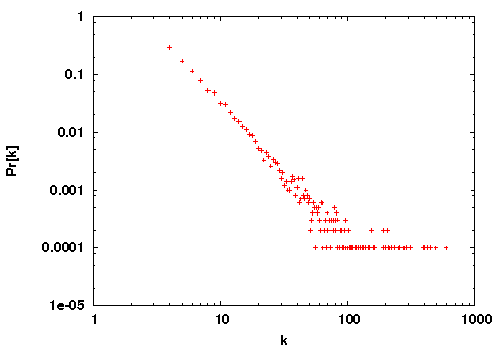}
\label{fig:subfig3}
}
\hspace{-5mm}
\subfigure[$m=5, \eta=1$]
{
\includegraphics[scale=0.33]{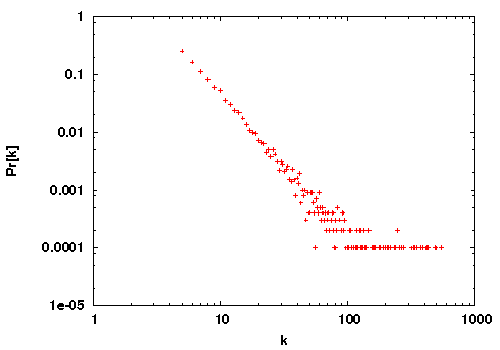}
\label{fig:subfig3}
}
\caption
{
\label{figure:degreedistribution2}
Comparison of degree distributions of four $\eta$ networks having different values of $m$.
}
\end{figure*}

As depicted in Equations \ref{eq:powerlaw1}, \ref{eq:gamma} and \ref{eq:finallowerbound},
another parameter affecting the $\eta$ networks is $m$.
In order to evaluate the influence of $m$, 
we fix $\eta$ to $1$ and $n$ to $10,000$,
and produce networks with different values for $m$: $2$, $3$, $4$ and $5$.
Figure \ref{figure:degreedistribution2} shows degree distributions of the produced networks.
As depicted in the figure, the degree distributions follow a power-law.
Similar to $\eta$, increasing $m$ slightly increases the exponent $\gamma$,
which is consistent with Equation \ref{eq:gamma}.
Figure \ref{figure:bound2} compares the clustering coefficient of the networks and the bound $B$ obtained in Equation \ref{eq:finallowerbound}.
In the produced networks, $n$ is $10,000$ and $\eta$ is $1$ and $m$ varies between $2$ to $5$.
It shows that by decreasing $B$, the clustering coefficient decreases as well
and as Equation \ref{eq:finallowerbound} says, increasing $m$, reduces $B$.
In Table \ref{table:dimacc2}, we describe the specifications of the networks.
By increasing $m$, both the clustering coefficient and the diameter decrease but the average degree increases.


\begin{table}
\caption{\label{table:dimacc2}
Diameter, clustering coefficient, and average degree of
networks produced by the $\eta$ model for different values of $m$.
$n$ is set to $10000$ and $\eta$ is set to $1$.
}
\centering
\begin{tabular}{l l l l}
\hline
$m$ & diameter & clustering coefficient & Avg. degree\\
\hline
$2$ & $4.34$ & $0.19$   & $5.826$ \\
$3$ & $3.88$ & $0.09$   & $7.804$ \\
$4$ & $3.32$ & $0.0638$ & $9.91$ \\
$5$ & $3.47$ & $0.05$   & $11.928$ \\
\hline
\end{tabular}
\end{table}

\section{Related work}
\label{sec:relatedwork}

In 
~\cite{conf:Aiello}, a power-law random graph model $P(\alpha, \beta)$ is proposed as follows:
let $y$ be the number of vertices with degree $x$. $P(\alpha,\beta)$ assigns uniform probability to all graphs with $y = e^{\alpha}/x^{\beta}$ where self loops are allowed.
The authors study the giant component and the evolution of random graphs in this model.
The authors of \cite{jrnl:Watts2} present a model to explain social network searchability
in terms of characteristics measured along a number of social dimensions.
Their model defines a class of searchable networks and a method for searching them.

Chung and Lu \cite{jrnl:Chung} consider a family of random graphs with a given expected degree sequence.
In this model each edge is selected independently with probability proportional to the product of the expected degrees of its endpoints.
They examine the distribution of the size of the connected components of the model.
Eubank et al. \cite{conf:Eubank} show that many basic characteristics of the social network for the city of Portland, Oregon, USA, are well-modeled by the random graph model of Chung and Lu.
They also give fast approximation algorithms for computing basic structural properties such as clustering coefficients and shortest paths distribution.

In \cite{jrnl:Moore}, the authors formulate models of the time evolution of the networks which acquire and lose vertices during the time.
They show that the model generates networks with power-law degree distributions with an exponent that diverges as the growth rate vanishes. 
In their models new vertices form links by preferential attachment, but the number of added vertices is equal to the number of deleted vertices.
In \cite{proceedings:Zhu} and \cite{jrnl:Zhu}, the authors study and analyze different properties like degree distribution, clustering coefficient, average path length and phase transition
of an evolving email network model.

Takemoto and Oosawa \cite{jrnl:Takemoto} propose a model for evolving networks by merging complete graphs (cliques) as building blocks.
The model shows power-law degree distribution, power-law clustering spectra and high average clustering coefficients independent of the size of network.
However, in most cases, real-world networks are formed in a different way:
they usually \textit{grow} during the time by obtaining new vertices,
rather than by merging complete graphs.

Serrano, Krioukov and Boguna \cite{jrnl:Serrano} prove that a class of hidden variable models with underlying metric spaces
are able to accurately reproduce the self-similarity properties in real-world networks.
They show that hidden metrics underlying these real-world networks can explain the topology of the networks.
Li and Maini \cite{jrnl:Li} propose an evolving network model which produces community structures.
The model is based on two mechanisms: the inner-community preferential attachment and inter-community preferential attachment.
However, while their theoretical (and numerical) simulations show that this network model has community structure,
they do not provide theoretical analysis for the clustering coefficient of the model.
The provided numerical simulations show that the clustering coefficient of the model
is still significantly lower than real-world networks.

Yang and Leskovec \cite{conf:Yang0} model the
global influence of a vertex on the rate of diffusion through the
network. They simulate the number of newly infected vertices as a function of which other vertices got infected in the
past. The same authors in \cite{conf:Yang}  examine several large scale social, collaboration and information networks and find that the community overlaps are more densely connected than the non-overlapping
parts. This is in contrast to the conventional wisdom that community overlaps are more sparsely connected than the communities themselves.
Kin and Leskovec \cite{conf:Kim0} propose the
Multiplicative Attribute Graphs (MAG) model which captures interactions between the
vertex attributes and the observed network structure.
In this model, the probability of having an edge between a pair of vertices depends on the individual attribute link formation affinities.
The same authors in \cite{conf:Kim} present a parameter estimation method for the MAG model which is based on variational expectation maximization.

\section{Conclusions}
\label{sec:conclusions}

In this paper, we proposed a new model, called the $\eta$ model, for describing transitivity relations in complex networks.
We theoretically analyzed the model and calculated a lower bound on the clustering coefficient of the model
which is independent of the network size 
and depends only on the model's parameters ($\eta$ and $m$).
We proved that the model satisfies important properties such as power-law degree distribution and the small-world property.
We also evaluated the model empirically and showed that it can precisely simulate real-world networks from different domains
with different specifications.

\bibliographystyle{plain}
\bibliography{references}

\begin{thebibliography}{33}
\providecommand{\natexlab}[1]{#1}
\providecommand{\url}[1]{\texttt{#1}}
\expandafter\ifx\csname urlstyle\endcsname\relax
  \providecommand{\doi}[1]{doi: #1}\else
  \providecommand{\doi}{doi: \begingroup \urlstyle{rm}\Url}\fi

\bibitem[Aiello et~al.(2000)Aiello, Chung, and Lu]{conf:Aiello}
William Aiello, Fan Chung, and Linyuan Lu.
\newblock A random graph model for massive graphs.
\newblock In \emph{Proceedings of the Thirty-second Annual ACM Symposium on
  Theory of Computing}, STOC '00, pages 171--180, 2000.

\bibitem[Albert and Barab\'asi(2002)]{jrnl:Albert}
R\'eka Albert and Albert-L\'aszl\'o Barab\'asi.
\newblock Statistical mechanics of complex networks.
\newblock \emph{Rev. Mod. Phys.}, 74\penalty0 (1):\penalty0 47--97, 2002.

\bibitem[Barabasi and Albert(1999)]{jrnl:Barabasi}
Albert-Laszlo Barabasi and Reka Albert.
\newblock Emergence of scaling in random networks.
\newblock \emph{Science}, 286\penalty0 (5439):\penalty0 509--512, 1999.

\bibitem[Barrat and Weigt(2000)]{jrnl:Barrat}
A.~Barrat and M.~Weigt.
\newblock On the properties of small-world network models.
\newblock \emph{EUROP.PHYS.J.B}, 13:\penalty0 547, 2000.

\bibitem[Cancho et~al.(2001)Cancho, Janssen, and Sole]{jrnl:Cancho}
R.~F. Cancho, C.~Janssen, and R.~V. Sole.
\newblock Topology of technology graphs: Small world patterns in electronic
  circuits.
\newblock \emph{Phys. Rev. E}, 64, 2001.

\bibitem[Chung and Lu()]{jrnl:Chung}
F.~Chung and L.~Lu.
\newblock \emph{Annals of Combinatorics}, page 125.

\bibitem[Cohen et~al.(1990)Cohen, Briand, and Newman]{jrnl:CohenFood}
Joel~E. Cohen, Frederic Briand, and Charles~M. Newman.
\newblock \emph{Community Food Webs: Data and Theory (Biomathematics)},
  volume~20.
\newblock Springer, 1990.

\bibitem[{Cohen} and {Havlin}(2003)]{jrnl:Cohen}
R.~{Cohen} and S.~{Havlin}.
\newblock {Scale-Free Networks Are Ultrasmall}.
\newblock \emph{Phys. Rev. Lett.}, 90:\penalty0 058701, 2003.

\bibitem[D.J.~Thouless and Palmer(1977)]{jrnl:TAP77}
P.W.~Anderson D.J.~Thouless and R.G. Palmer.
\newblock A solution to a solvable model of a spin glass.
\newblock \emph{Philosophical Magazine}, page 35:593, 1977.

\bibitem[Erdős and Rényi(1960)]{jrnl:Erdos}
P.~Erdős and A~Rényi.
\newblock On the evolution of random graphs.
\newblock pages 17--61, 1960.

\bibitem[Eubank et~al.(2004)Eubank, Kumar, Marathe, Srinivasan, and
  Wang]{conf:Eubank}
Stephen Eubank, V.~S.~Anil Kumar, Madhav~V. Marathe, Aravind Srinivasan, and
  Nan Wang.
\newblock Structural and algorithmic aspects of massive social networks.
\newblock In \emph{Proceedings of the Fifteenth Annual ACM-SIAM Symposium on
  Discrete Algorithms}, SODA '04, pages 718--727, 2004.

\bibitem[Gr{\"o}tschel et~al.(2001)Gr{\"o}tschel, Krumke, and
  Rambau]{jrnl:GKR01}
Martin Gr{\"o}tschel, Sven~O. Krumke, and J{\"o}rg Rambau, editors.
\newblock \emph{Online Optimization of Large Scale Systems}.
\newblock Springer, 2001.

\bibitem[Hofmann and Buhmann(1997)]{jrnl:HB97}
Thomas Hofmann and Joachim~M. Buhmann.
\newblock Pairwise data clustering by deterministic annealing.
\newblock \emph{IEEE Trans. Pattern Anal. Mach. Intell.}, 19\penalty0
  (1):\penalty0 1--14, 1997.
\newblock URL
  \url{http://dblp.uni-trier.de/db/journals/pami/pami19.html#HofmannB97}.

\bibitem[Huxham et~al.(1996)Huxham, Beaney, and Raffaelli]{jrnl:Huxham}
M.~Huxham, S.~Beaney, and D.~Raffaelli.
\newblock Do parasites reduce the chances of triangulation in a real food web?
\newblock \emph{Oikos}, 76:\penalty0 284--300, 1996.

\bibitem[Kim and Leskovec(2011)]{conf:Kim}
Myunghwan Kim and Jure Leskovec.
\newblock Modeling social networks with node attributes using the
  multiplicative attribute graph model.
\newblock pages 400--409, 2011.

\bibitem[Kim and Leskovec(2012)]{conf:Kim0}
Myunghwan Kim and Jure Leskovec.
\newblock Multiplicative attribute graph model of real-world networks.
\newblock \emph{Internet Mathematics}, 8\penalty0 (1-2):\penalty0 113--160,
  2012.

\bibitem[Knuth(1993)]{book:Knuth}
Donald~E. Knuth.
\newblock \emph{The Stanford GraphBase - a platform for combinatorial
  computing.}
\newblock ACM, 1993.

\bibitem[Leskovec et~al.(2008)Leskovec, Backstrom, Kumar, and
  Tomkins]{conf:Leskovec}
Jure Leskovec, Lars Backstrom, Ravi Kumar, and Andrew Tomkins.
\newblock Microscopic evolution of social networks.
\newblock In \emph{KDD '08: Proceeding of the 14th ACM SIGKDD international
  conference on Knowledge discovery and data mining}, pages 462--470, 2008.

\bibitem[Li and Maini(2005)]{jrnl:Li}
C.~Li and P.~K. Maini.
\newblock An evolving network model with community structure.
\newblock \emph{Journal of Physics A: Mathematical and General}, 38:\penalty0
  9741--9749, 2005.

\bibitem[Moore et~al.(2006)Moore, Ghoshal, and Newman]{jrnl:Moore}
Cristopher Moore, Gourab Ghoshal, and M.~E.~J. Newman.
\newblock Exact solutions for models of evolving networks with addition and
  deletion of nodes.
\newblock 74\penalty0 (036121), 2006.

\bibitem[Newman et~al.(2001)Newman, Strogatz, and Watts]{jrnl:Newman2}
M.~E.~J. Newman, S.~H. Strogatz, and D.~J. Watts.
\newblock Random graphs with arbitrary degree distributions and their
  applications.
\newblock \emph{Physical Review E}, 64\penalty0 (2):\penalty0 026118, 2001.

\bibitem[Newman et~al.(2002)Newman, Forrest, and Balthrop]{jrnl:NewmanEmail}
M.~E.~J. Newman, Stephanie Forrest, and Justin Balthrop.
\newblock Email networks and the spread of computer viruses.
\newblock \emph{Phys. Rev. E}, 66\penalty0 (3):\penalty0 035101, 2002.

\bibitem[Puzicha et~al.(2000)Puzicha, Hofmann, and Buhmann]{jrnl:PHB00}
Jan Puzicha, Thomas Hofmann, and Joachim~M. Buhmann.
\newblock A theory of proximity based clustering: structure detection by
  optimization.
\newblock \emph{Pattern Recognition}, 33\penalty0 (4):\penalty0 617--634, 2000.
\newblock \doi{10.1016/S0031-3203(99)00076-X}.
\newblock URL \url{http://dx.doi.org/10.1016/S0031-3203(99)00076-X}.

\bibitem[Serrano et~al.(2008)Serrano, Krioukov, and Boguna]{jrnl:Serrano}
M~Angeles Serrano, Dimitri Krioukov, and Marian Boguna.
\newblock Self-similarity of complex networks and hidden metric spaces.
\newblock \emph{Phys. Rev. Lett.}, 100\penalty0 (078701), February 2008.

\bibitem[Takemoto and Oosawa(2005)]{jrnl:Takemoto}
Kazuhiro Takemoto and Chikoo Oosawa.
\newblock Evolving networks by merging cliques.
\newblock \emph{Phys. Rev. E}, 742, 2005.

\bibitem[Thomas and Finney(1996)]{book:thomas}
G.B. Thomas and R.L. Finney.
\newblock \emph{Calculus and analytic geometry}.
\newblock Calculus and Analytic Geometry. Addison-Wesley, 1996.

\bibitem[Watts and Strogatz(1998)]{jrnl:Watts}
D.J. Watts and S.H. Strogatz.
\newblock Collective dynamics of 'small-world' networks.
\newblock \emph{Nature}, \penalty0 (393):\penalty0 409--410, 1998.

\bibitem[Watts et~al.(2002)Watts, Dodds, and Newman]{jrnl:Watts2}
D.J. Watts, P.S. Dodds, and M.E.J. Newman.
\newblock Identity and search in social networks.
\newblock \emph{Science}, 296:\penalty0 1302, 2002.

\bibitem[White et~al.(1986)White, Southgate, Thomson, and Brenner]{jrnl:White}
J.G. White, E.~Southgate, J.~N. Thomson, and S.~Brenner.
\newblock The structure of the nervous system of the nematode c. elegans.
\newblock \emph{Philosophical transactions Royal Society London}, 314:\penalty0
  1--340, 1986.

\bibitem[Yang and Leskovec(2010)]{conf:Yang0}
Jaewon Yang and Jure Leskovec.
\newblock Modeling information diffusion in implicit networks.
\newblock In \emph{Proceedings of the 2010 IEEE International Conference on
  Data Mining}, ICDM '10, pages 599--608. IEEE Computer Society, 2010.

\bibitem[Yang and Leskovec(2012)]{conf:Yang}
Jaewon Yang and Jure Leskovec.
\newblock Community-affiliation graph model for overlapping network community
  detection.
\newblock In \emph{Proceedings of the 2010 IEEE International Conference on
  Data Mining}, ICDM '12, pages 1170--1175. IEEE Computer Society, 2012.

\bibitem[Zhu and Kuh(2006)]{proceedings:Zhu}
Chaopin Zhu and Anthony Kuh.
\newblock On randomly evolving email networks.
\newblock In \emph{39th Annual Conference on Information Sciences and Systems},
  2006.

\bibitem[Zhu et~al.(2006)Zhu, Kuh, Wang, and De~Wilde]{jrnl:Zhu}
Chaopin Zhu, Anthony Kuh, Juan Wang, and Philippe De~Wilde.
\newblock Analysis of an evolving email network.
\newblock \emph{Phys. Rev. E}, 74:\penalty0 046--109, 2006.

\end{thebibliography}

\end{document}